\documentclass[12pt,preprint]{aastex}

\newcommand{\simgt}{\,\rlap{\lower 3.5 pt \hbox{$\mathchar \sim$}} \raise
1pt \hbox {$>$}\,}
\newcommand{\simlt}{\,\rlap{\lower 3.5 pt \hbox{$\mathchar \sim$}} \raise
1pt \hbox {$<$}\,}

\shorttitle{Cluster temperature from the thermal 
Sunyaev-Zel'dovich effect}
\shortauthors{Hansen, Pastor \& Semikoz}

\begin{document}


\title{First measurement of cluster temperature\\ using the thermal 
Sunyaev-Zel'dovich effect}


\author{Steen H. Hansen}
\affil{University of Oxford, Astrophysics, Keble Road, 
Oxford, OX1 3RH, United Kingdom}
\email{hansen@astro.ox.ac.uk}

\author{Sergio Pastor and Dmitry V. Semikoz\altaffilmark{1}}
\affil{Max-Planck-Institut f\"ur Physik (Werner-Heisenberg-Institut)
F\"ohringer Ring 6, 80805 M\"unchen, Germany}
\email{pastor@mppmu.mpg.de, semikoz@mppmu.mpg.de}


\altaffiltext{1}{Institute of Nuclear Research of the Russian Academy
of Sciences, 60th October Anniversary Prospect 7a, Moscow 117312,
Russia}


\begin{abstract}
We discuss a new method of finding the cluster temperatures which is
independent of distance and therefore very useful for distant
clusters.  The hot gas of electrons in clusters of galaxies scatters
and distorts the cosmic microwave background radiation in a well
determined way. This Sunyaev-Zel'dovich (SZ) effect is a useful tool
for extracting information about clusters such as their peculiar
radial velocity and optical depth. Here we show how the temperature of
the cluster can be inferred from the SZ effect, in principle without
use of X-ray data. We use recent millimetre observation of Abell 2163
to determine for the first time a cluster temperature using SZ
observations only. The result $T_e = 26^{+34}_{-19}$ keV at 68\%
confidence level (at 95\% c.l. we find $T>1.5$ keV) is in
reasonable agreement with the X-ray results, $T_e
=12.4^{+2.8}_{-1.9}$ keV.
\end{abstract}




The interaction of the Cosmic Microwave Background Radiation (CMBR)
photons with the free electrons in the ionized gas of clusters of
galaxies produces the thermal Sunyaev-Zel'dovich effect \citep{sz}.
This distortion imprinted on the spectrum of the CMBR is independent
of redshift and can be used, in combination with other observations
like X-ray measurements of the clusters, to extract important
cosmological parameters such as the Hubble constant or the total mass
density of the universe~\citep{Birkinshaw99,Carlstrom01}.  Nowadays
there exist accurate measurements of the SZ effect in several
clusters, and very recently the measurements of the SZ effects at high
frequencies produced the first multi-frequency spectra, in the cases of
the Abell 2163 cluster \citep{LaRoque01}, and the Coma
cluster~\citep{Coma}.

The intensity change of the CMBR is proportional to the Comptonization
parameter,
\begin{equation}
y_c = \int dl~ \frac{T_e}{m_e} ~ n_e \sigma_{\rm Th} ~,
\end{equation}
where $T_e$ is the average temperature of the electron gas, $m_e$ the
electron mass, $n_e$ the electron number density, $\sigma_{\rm Th}$
the Thomson scattering cross section, and the integral is calculated
along the line of sight through the cluster. We use units for which
$k_B=\hbar=c=1$.  For an intra-cluster gas which can be assumed
isothermal one has $y_c=\tau \, T_e/m_e$, where $\tau$ is the optical
depth. The intensity change is given by
\begin{equation}
\Delta I_{\rm T} = 
I_0~ y_c~ \left( \frac{x^4e^x}{(e^x-1)^2}\left[
\frac{x(e^x+1)}{e^x-1}-4\right]~ +\delta f(x,T_e)\right)~,
\label{thermal}
\end{equation}
where $x=\nu/T_{\rm CMB}$ is the dimensionless frequency ($T_{\rm CMB}
=2.725 ~{\rm K}$), and $I_0=T_{\rm CMB}^3/(2\pi^2)$. The intensity
change is independent of the temperature for non-relativistic
electrons, a limit which is valid for small frequencies ($\nu\simlt
100$ GHz), but for high frequencies it must either be
corrected~\citep{rephaeli95} with $\delta f(x,T_e)$ by using an expansion
in $\theta_e=T_e/m_e$ \citep{stebbins97,Challinor98,ItohI,ItohV} or
calculated exactly \citep{dolgov00}. A useful analytic fitting formula
valid for $\theta_e \leq 0.05$ ($T_e \leq 25.5$ keV) has recently been
given by \citet{ItohIV}. The corrections for larger temperatures must
be found with the exact method of \citet{dolgov00}.

%In particular these relativistic corrections are important for 

\begin{figure}[htb!]
\plotone{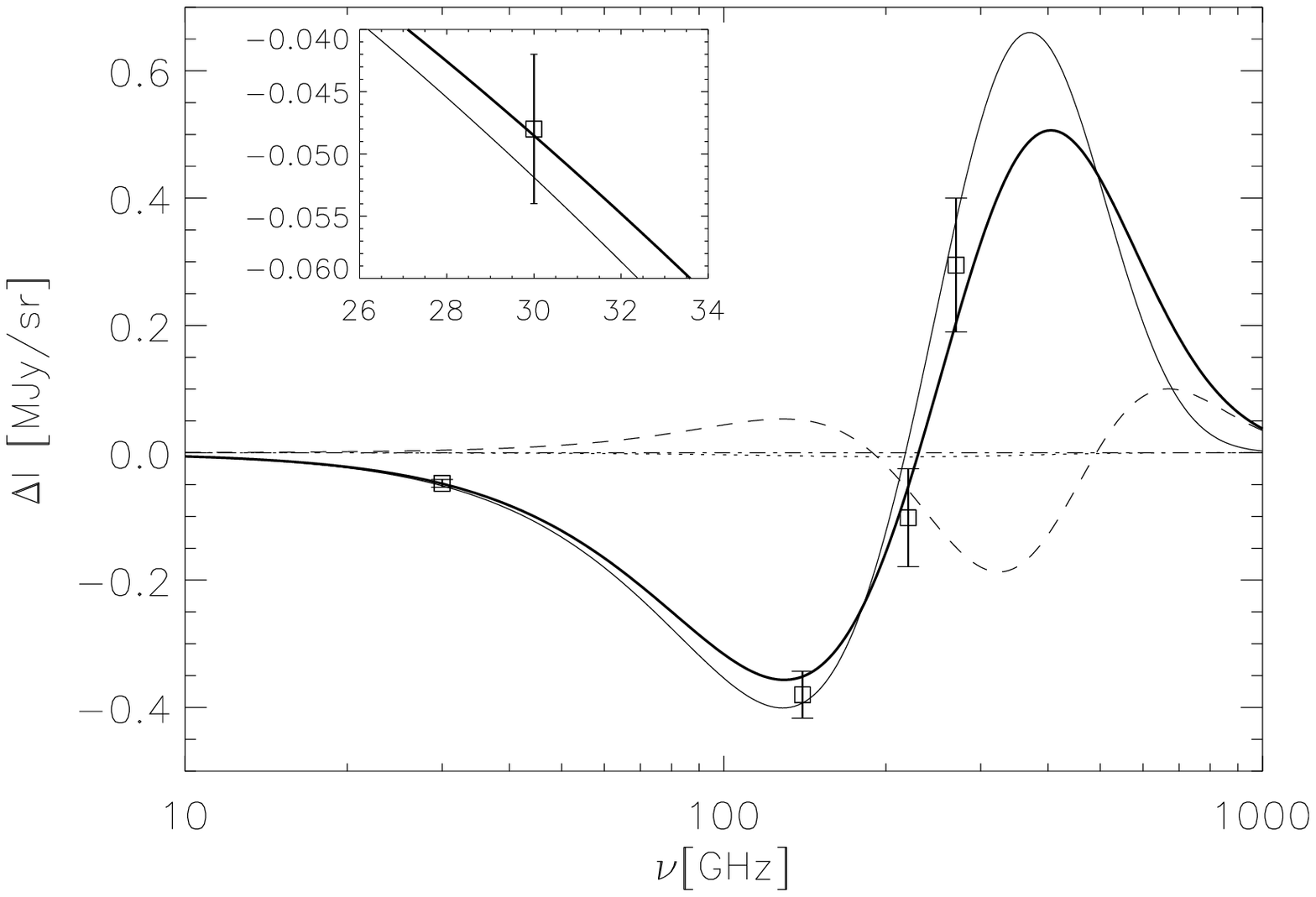}
\caption{Observational data for the SZ effect in Abell 2163 for the
30, 140, 220 and 270 GHz frequency bands from
\citet{LaRoque01} and \citet{SuZIE}, together with theoretical predictions
for the parameter choice $T_e=26$ keV, $\tau = 7.2 \times 10^{-3}$,
and $v_p = 165 ~{\rm km}/{\rm s}$. The thick solid line is the total
SZ effect, the thin solid line is the thermal effect with vanishing
electron temperature, the dashed line is the contribution of
relativistic corrections (the $\delta f(x,T_e)$ term in
Eq.~(\ref{thermal})), and the dotted line is the kinematic
effect. \label{fig1}}
\end{figure}

An additional distortion is caused by the relative motion of the
cluster with respect to the CMBR rest frame, an effect known as the
kinematic SZ effect \citep{sz80}, which is identical to a change in the
temperature of the CMBR.  The corresponding intensity change, for a
peculiar velocity $v_p$, is
\begin{equation}
\Delta I_{\rm K} = -  I_0 \tau \frac{v_p}{c} \frac{x^4 e^x}{(e^x-1)^2}~, 
\label{kinematic}
\end{equation}
We do not include the temperature dependence of this
kinetic distortion \citep{sazonov,ItohII}, since it only gives a second order
correction to the main effect. The total intensity change is just
the sum of the two SZ effects,
\begin{equation}
\Delta I_{\rm total} = \Delta I_{\rm T}+\Delta I_{\rm K}~.
\label{total}
\end{equation}
The different contributions to the SZ distortion are presented in
Fig.~\ref{fig1}, where the total intensity change in Eq.~(\ref{total})
is compared to the thermal effect with vanishing temperature (the
first term in Eq.~(\ref{thermal})). The kinematic SZ effect and the
contribution of the relativistic corrections (the second term in
Eq.~(\ref{thermal})) are presented for the parameter choice $T_e=26$
keV, $\tau = 7.2 \times 10^{-3}$, and $v_p = 165 ~{\rm km}/{\rm s}$.
One can see the importance of the relativistic
corrections for such a high temperature on figure \ref{fig1}.

Now, the theoretical prediction can be compared to observations in
order to extract the unknown values of the optical depth $\tau$,
peculiar velocity $v_p$, and electron temperature $T_e$. The value of
$y_c$ is basically derived from the SZ measurement at low frequencies,
where the temperature corrections are negligible, while the kinematic
effect is maximal at the cross-over frequency, where the thermal SZ
effect vanishes ($\nu_0=217$~GHz when $T_e \rightarrow 0$).  Finally
the data at higher frequencies may be used to find the non-trivial
dependence of the electron temperature from the relativistic
corrections in the second term of Eq.~(\ref{thermal}).

Thus, with SZ observations at several different frequencies one can
obtain the value of the electron temperature, independent of X-ray
measurements. It is then clear that the error-bar of the temperature
determination is mainly determined by the error-bars on the
observations near $300$ GHz.  A forecast for the ability of Planck and
FIRST to extract cluster temperatures using this technique was
ana\-lysed in \citet{pgb98}.

A different technique to extract the temperature of a cluster uses the
gas mass fraction extracted from a sample of clusters with both SZ and
X-ray observations. Assuming that this gas mass fraction is universal,
one can thereby extract the cluster temperature \citep{joy2001}. This
method is complementary to ours, since the one we are describing does
not need this assumption.

\begin{figure}[htb!]
\plotone{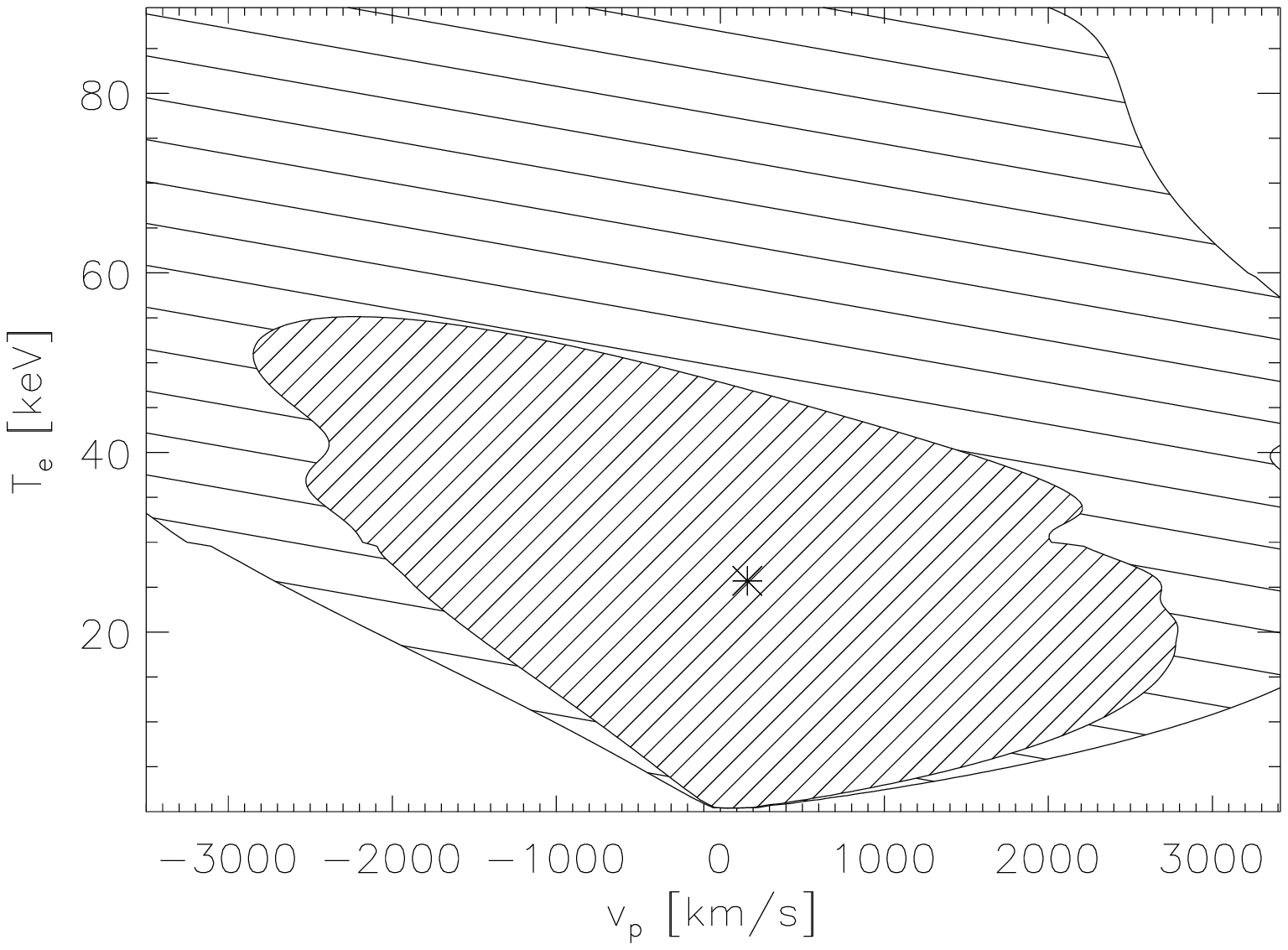}
\caption{The marginalized likelihood in ($T_e,v_p$) plane.  The
densely hatched region indicates $\Delta \chi^2=2.3$ ($\sim 1
\sigma$), and the less dense region is for $\Delta \chi^2=4.6$ ($\sim
2 \sigma$).  Our resulting central value is shown by a
star.\label{fig2}}
\end{figure}
Previous analyses of SZ cluster data have shown that it is possible to
put bounds on the values of the Comptonization parameter and the
peculiar velocity~\citep{LaRoque01}.  Here we extend these analyses
and derive the electron temperature using exclusively the existing SZ
measurements of the galaxy cluster Abell 2163 \citep{LaRoque01,SuZIE},
one of the hottest and most luminous of known clusters.  We benefit
from the remarkable progress in SZ observations over recent years,
and use SZ data of Abell 2163 at redshift $z=0.2$ from the
Berkeley-Illinois-Maryland Association (BIMA) and Owens Valley Radio
Observatory (OVRO) millimeter interferometers \citep{LaRoque01} at a
frequency of 30 GHz, and the dust-corrected data from the
Sunyaev-Zel'dovich Infrared Experiment (SuZIE) at 140, 220 and 270 GHz
\citep{SuZIE} to perform a likelihood analysis varying the three
parameters: $\tau$, $v_p$ and $T_e$, allowed to vary in the ranges $
0.001< \tau < 0.4$, $ -3500 < v_p < 3500 $ km/s, and $ 0 < T_e <90$
keV.  The SZ distortions are calculated with the exact method
described in \citet{dolgov00}, which allows us to calculate the SZ
distortion accurately for any temperature.  The resulting bounds are
marginalization over all other parameters, and the $68\%$ confidence
level (c.l.) is thus the integrated likelihood. The fairly
non-elliptical likelihood contours are exemplified in Fig.~\ref{fig2},
where indicative $\Delta \chi^2=2.3$ and $4.6$ are shown in the
($T_e,v_p$) plane. The resulting central value is shown with a star.
The results of this combined analysis are: $v_p = 165^{+1350}_{-1700}$
km/s, $\tau = 0.0072^{+0.016}_{-0.0015}$, and $T_e = 26^{+34}_{-19}$
keV. At 95\% c.l. we find $T>1.5$ keV, and $-5000 > v_p > 4000$ km/s.
The value of the electron temperature is in reasonable
agreement with that extracted from the X-ray data: $T_e
=12.4^{+2.8}_{-1.9}$ keV \citep{SuZIE,Holzapfelb}. The large error-bar
on the SZ temperature determination depends strongly on the large
error-bar of the $270$ GHz observation, which will be significantly
improved with future observations. If we artificially reduce all the
observational error-bars by a factor of 2 (3), then the resulting
error-bar on the temperature improves by almost a factor of 3 (5).
The mass of the cluster is directly related to the electron
temperature through~\citep{Finoguenov} $M \approx 4.2 \cdot 10^{13}
(T/\mbox{keV})^{1.48} M_\odot$, which for our central value of $T=26$
keV corresponds to $M\approx5 \cdot 10^{13} M_\odot$.

The main application of temperature determination using SZ
observations is for clusters at large distances. This is because the
redshift independence of SZ makes it more powerful for very distant
clusters than X-ray observations for which the X-ray flux diminishes
rapidly~\citep{korolyov}.  Knowledge of the temperatures of distant
clusters is crucial for the determination of cluster abundance and
understanding their evolution.  A mea\-surement of the cluster
abundance at large redshift will allow precise determination of
cosmological parameters~\citep{henry}, and in combination with other
observations can be used to constrain primordial
non-Gaussianity~\citep{nongaus}.  Also an inhomogeneous intergalactic
gas can be studied with the expected improved SZ observations in the
near future, using either bolometer or interferometer-based surveys.
Furthermore, SZ temperature determination can be used as complementary
information in merging clusters with multiple temperatures or
non-thermal electrons, where the interpretation of X-ray results may
be particularly difficult.

\acknowledgments

It is a pleasure to thank G.~Bryan, J.~Silk, R.~A.~Sunyaev and J.~Wall
for stimulating discussions. In Munich, this work was partly supported
by the Deut\-sche For\-schungs\-ge\-mein\-schaft under grant No.\ SFB
375 and the ESF network Neutrino Astrophysics. SHH and SP are
supported by Marie Curie Fellowships of the European Commission under
contracts HPMFCT-2000-00607 and HPMFCT-2000-00445.

\end{document}